\documentclass[preprint]{elsarticle}

\usepackage{lineno,hyperref}
\modulolinenumbers[5]
\usepackage{caption} 
\journal{Journal of \LaTeX\ Templates}

\begin{document}

\begin{frontmatter}

\title{Modeling the Generic Breakthrough Curve for Adsorption Process}
%\tnotetext[mytitlenote]{Fully documented templates are available in the elsarticle package on \href{http://www.ctan.org/tex-archive/macros/latex/contrib/elsarticle}{CTAN}.}

%% Group authors per affiliation:
%\author{Saikat Roy\fnref{myfootnote}}
%\address{Radarweg 29, Amsterdam}
%\fntext[myfootnote]{Since 1880.}

%% or include affiliations in footnotes:
\author[mymainaddress]{Saikat Roy\corref{mycorrespondingauthor}}
\cortext[mycorrespondingauthor]{Corresponding author}
\ead{saikat.roy1988@gmail.com}

\author[mysecondaryaddress]{Arun S. Moharir}

\address[mymainaddress]{Dept of Chemical Physics, The Weizmann Institute of Science, Israel}
\address[mysecondaryaddress]{Indian Institute of Technology-Bombay, India}

\begin{abstract}
This work is aimed at understanding the basic principles of adsorption process in great details as adsorptive separation process has broad applications in the industry. To this end, a simple mathematical model has been used to describe transient fixed bed physical adsorption process. Governing equations are solved numerically to obtain breakthrough curves for single component and multi-component monolayer adsorption. Desorption of a saturated bed by an inert fluid is also considered. A full parametric study is performed to analyze the effects of different parameters such as bed length, velocity, diffusivity, particle radius and isotherm properties on the nature of the breakthrough curve. Analysis of these results led to the development of the generic breakthrough curve for a single component monolayer adsorption which will enable us to tell the nature of breakthrough curve for different process parameters without recourse to the numerical simulation or experiment. Thus this study will be  of great interest in the industrial separation process.
\end{abstract}

\begin{keyword}
\texttt{Adsorption}\sep \texttt{Generic Breakthrough curve} \sep \texttt{Mathematical model} \sep \texttt{Linear driving force}
%\MSC[2010] 00-01\sep  99-00
\end{keyword}

\end{frontmatter}

%\linenumbers

\section{Introduction}

Every solid surface has a discontinuous structure where unsaturated forces act. When the
solid is exposed to a fluid, the fluid molecules get attached to the surface by forming bonds.
This phenomenon is known as adsorption. In the past two decades adsorption has become a
key process for fluid separation in the chemical and petrochemical industries following the
developments of new adsorbents and new process cycles. Invention of synthetic zeolites or
molecular sieve carbon marked the beginning of a new era in separation process by
adsorption. These types of adsorbents preferentially adsorb nitrogen over oxygen by an
approximate factor of three. So this can be a viable alternative to the conventional cryogenic
air separation process with the help of an efficient process cycle. Regeneration of the process
is also important for the reuse of the bed and recovery of the adsorbed fluid. Regeneration by
heating is not so efficient as heating-cooling cycle is time consuming. So this led to the
development of efficient pressure swing adsorption cycles \cite{jain2003heuristic,chahbani2000mass}. This is a very energy efficient
cycle. Regeneration can take place within minutes or seconds by lowering the pressure. This allows higher throughput of gas to be separated. Cryogenic air separation requires liquefaction
of air followed by distillation. Although this is the most frequently used process for large
scale air separation, it is not so energy efficient as a low temperature and a high pressure is
required for liquefaction. Following criteria describe where adsorption can be used over
distillation. 
\begin{itemize}
\item The relative volatility between the key components to be separated is in the order of
1.2 to 1.5 or less. Separation of isomers is an example. Adsorption separation factor
for separation of isomers by zeolite is infinite.

\item The two groups of components to be separated have overlapping boiling point. This
will require several distillation columns for the desired separation. The said
separation can efficiently be done by adsorption if the two groups contain chemically
or geometrically dissimilar molecules.

\item Major cost of a pressure swing adsorption is the compressor costs. If the feed is
available at an elevated pressure, cost will be drastically reduced.
\end{itemize}

It has been observed that it is better to opt for the adsorption separation over distillation for
small to medium throughput and when high purity products are not required. As the
development of new adsorbent and modification of process cycles takes place, pressure
swing adsorption will compete with distillation at high throughput\cite{ruthven1984principles}.
Based on the method of adsorbent regeneration, adsorptive separation can be categorized
in the following process cycles
\begin{itemize}
\item Temperature swing adsorption (TSA).
\item Pressure swing adsorption (PSA).
\item Inert purge cycle.
\item Displacement purge cycle.
\end{itemize}
Adsorptive separation has been used extensively in industry for different purposes like drying
of cracked gas, ethylene; n-Paraffin recovery from naphtha and kerosene; aromatic
separation; oxygen and nitrogen production from air; high purity hydrogen production from
steam reformer products, solid liquid separation and many more. Zeolite, activated carbon, molecular sieve carbon, activated alumina
are generally used as adsorbents  in industry for the aforementioned purposes.
Most of the adsorptive separation takes place in a fixed bed packed with adsorbent particles.
In order to find out the fluid phase adsorbate concentration profile within the bed and also to
find out the effluent concentration history i.e. breakthrough curve a model is required to
account for the physical processes happening in the bed. In this report a mathematical model
\cite{ruthven1984principles,yang2013gas}has been used with proper boundary and initial conditions to
predict the breakthrough nature for single component monolayer adsorption onto an
adsorbent. It has also been extended for the multi-component case. And also the breakthrough
curve for desorption by an inert is predicted through the same model with different boundary
and initial conditions. Although many experimental and simulation studies exist in the literature to find the breakthrough nature for the adsorption process \cite{casas2012fixed,ribeiro2008adsorption,poursaeidesfahani2019prediction,serna2010modeling,al2018breakthrough}  , but most of the studies are confined to specific set of cases. In this work, we propose a model to find out a generic breakthrough curve
which will predict breakthrough nature of adsorption for different values of process variables without
recourse to the numerical simulation or experiment.

\section{Breakthrough nature for single component adsorption}

\subsection{Mathematical modelling}

In the present work following assumptions are made to model the fixed bed adsorption
process:
\begin{enumerate}
\item The system operates under isothermal condition.
\item Negligible pressure drop through the adsorbent bed.
\item Velocity is assumed to be constant throughout the bed.
\item Langmuir isotherm is valid for the system.
\item Ideal plug flow is assumed; i.e there is no axial or radial dispersion.
\item The mass transfer rate is represented by a linear driving force expression \cite{alpay1992linear}.
\item The bed is clean initially.
\end{enumerate}

Based on the above said assumptions, for the control volume $A\times dz$ and for the limit $z\rightarrow0$, the net rate of accumulation is given as,

\begin{equation}
u \frac{\partial c}{\partial z}+\frac{\partial c}{\partial t}+\frac{1-\epsilon}{\epsilon}\rho_p\frac{\partial q}{\partial t}=0
\label{massbalance}
\end{equation}
whereas $c$ is the concentration in fluid phase ($kg/m^3$), $\epsilon$ is the bed porosity , $\rho_p$ is adsorbent density ($kg/m^3$), $q$ is adsorbed phase concentration (kg adsorbed/ kg of adsorbent).

The mass transfer kinetics is modeled using the LDF, Linear driving force approximations, based on the simplifications of Fick's second law of diffusion,

\begin{equation}
\frac{\partial q}{\partial t}=\frac{15D_e}{R_p^2}\left(q^{*}-q\right)
\label{masstransfer}
\end{equation}

whereas $D_e$ is intracrystalline diffusivity, $R_p$ is adsorbent particle radius, $q^{*}$ is equilibrium concentration of adsorbed phase (kg adsorbed/ kg of adsorbent).
The adsorption isotherm is described by Langmuir,
\begin{equation}
q^*=\frac{q_mbc}{1+bc}
\label{isotherm}
\end{equation}
whereas $q_m$ is maximum adsorption capacity (kg adsorbed / kg of adsorbent) and $b$ is Langmuir isotherm constant ($m^3/kg$).
The following initial conditions are considered,

\begin{eqnarray}
\nonumber c=c_o &&  z=0, t=0 \\ \nonumber
q=0 &&    0<z\leq L, t=0 \\ 
c=0 &&   0<z\leq L, t=0
\label{icssingle}
\end{eqnarray}

The boundary conditions are given as ,
\begin{eqnarray}
c=c_o &&  z=0,t>0
\label{bcssingle}
\end{eqnarray}

We now substitue Eqn.~ \ref{masstransfer} in Eqn.~\ref{massbalance} and on using expression for isotherm(Eqn.~\ref{isotherm}) we get the following coupled PDEs,

\begin{equation}
u \frac{\partial c}{\partial z}+\frac{\partial c}{\partial t}+\frac{1-\epsilon}{\epsilon}\rho_p\frac{15D_e}{R_p^2}\left(\frac{q_mbc}{1+bc}-q\right)=0
\label{massbalancemod}
\end{equation}

\begin{equation}
\frac{\partial q}{\partial t}=\frac{15D_e}{R_p^2}\left(\frac{q_mbc}{1+bc}-q\right)
\label{masstransfermod}
\end{equation}

\paragraph{Simulation technique:}

Preceding set of partial differential equations is initially discretized by finite difference
method to obtain a set of algebraic equations. Then it is solved by explicit Euler method to get
the concentration profile for both in fluid and solid phase as a function of time and space and
also to obtain outlet concentration profile with respect to time. A mathematical algorithm to
solve these coupled equations is developed and implemented into a computer program using
MATLAB software.
\begin{table}
\caption*{Table 1}
\begin{center}
  \begin{tabular}{ |c|c|c|}
    \hline
    Parameter & Value \\ \hline
    $L$,Bed length,m & $0.5$ \\ \hline
    $\epsilon$ & $0.4$ \\ \hline
    $u$, Velocity ,$m/s$ & $0.01$ \\ \hline
    $\frac{15D_e}{R_p^2}$ & 0.5 \\ \hline
    $b$ & $0.3$ \\ \hline
    $q_m$ & $0.04$ \\ \hline
    $c_o$, inlet conc. of the adsorbate in the feed , $kg/m^3$ & $1$ \\ \hline
    $\rho_p$ & $1000$\\ \hline
  
  \end{tabular}
 
  \label{parameterz}
\end{center}
\end{table}

\begin{figure}
\includegraphics[scale=.35]{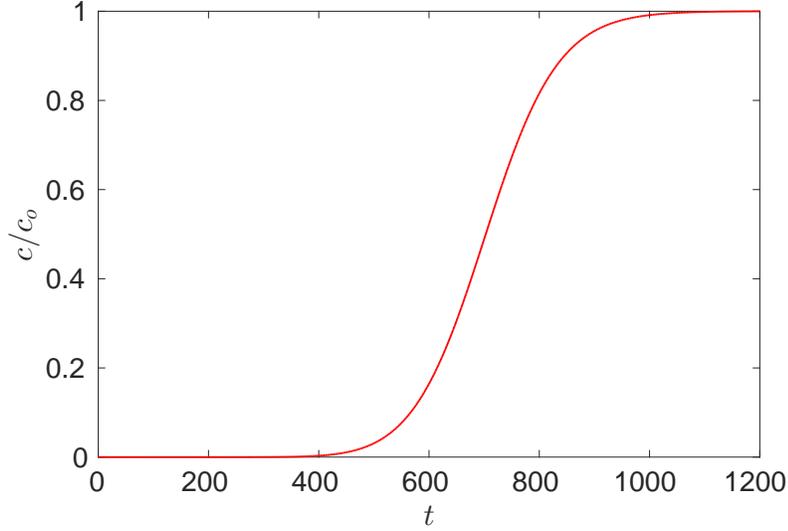}
\caption{Breakthrough curve for single component adsorption}
\label{single}
\end{figure}

\begin{figure}
\includegraphics[scale=.35]{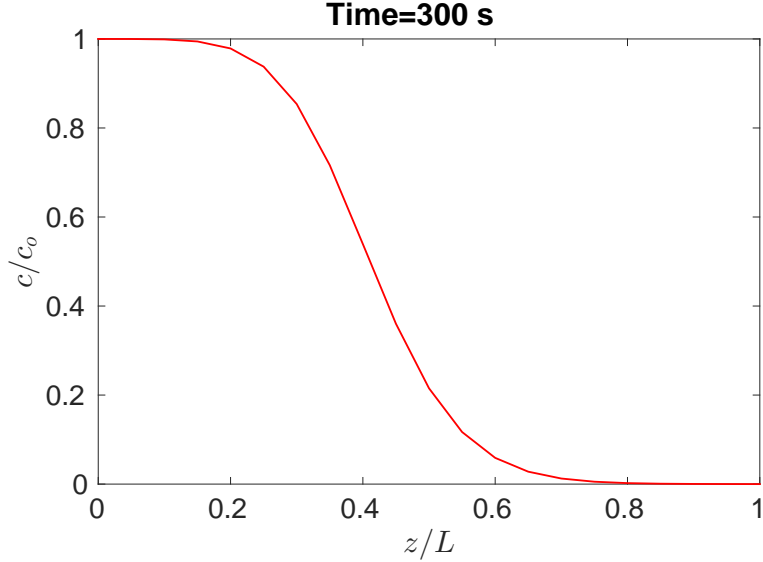}
\caption{Fluid phase concentration profile in the bed at a particular time}
\label{singlez}
\end{figure}

\paragraph{Results:}
 The breakthrough curve obtained from the simulation is shown in Fig. \ref{single}. The values of the parameter used in the simulation are given in Table 1. From the breakthrough curve we can find out at what time $ 90\% $ or $ 100\% $ breakthrough is taking place and accordingly feed will be switched to another bed until the saturated bed gets regenerated. Fluid phase concentration profile of adsorbate within the bed at a particular time is also shown in Fig. \ref{singlez}

\section{Single component desorption by an inert fluid}

\paragraph{Mathematical modeling}

Governing equations for desorption is same as that of adsorption. Only initial conditions and
boundary conditions are different. So the same model described earlier is used. Modified boundary and initial conditions are given as follows,

Boundary condition:

\begin{eqnarray}
c=0 && z=0,t>0
\end{eqnarray}

Initial conditions:

\begin{eqnarray}
\nonumber q=q_o^{*} && 0<z\leq L,t=0 \\ 
c=c_o && 0<z\leq L,t=0
\end{eqnarray}

\paragraph{Results}

\begin{figure}
\includegraphics[scale=.35]{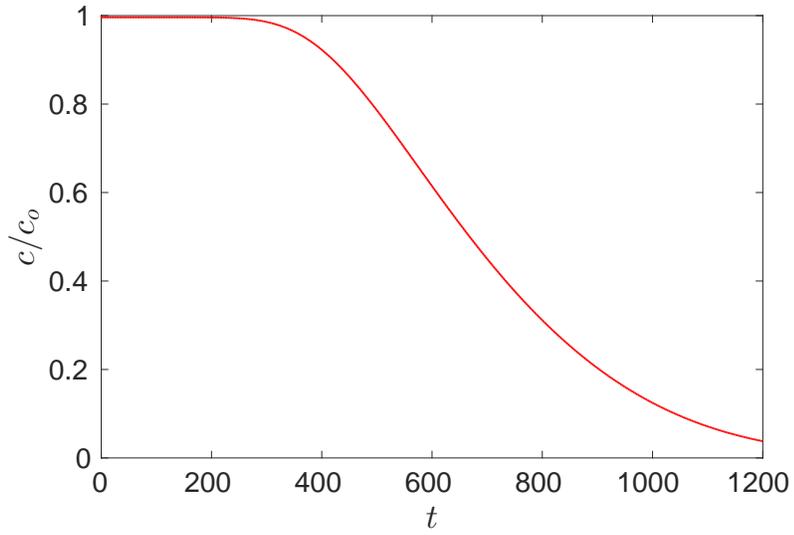}
\caption{Desorption breakthrough curve for single component}
\label{singlezdesorp}
\end{figure}

\begin{figure}
\includegraphics[scale=.35]{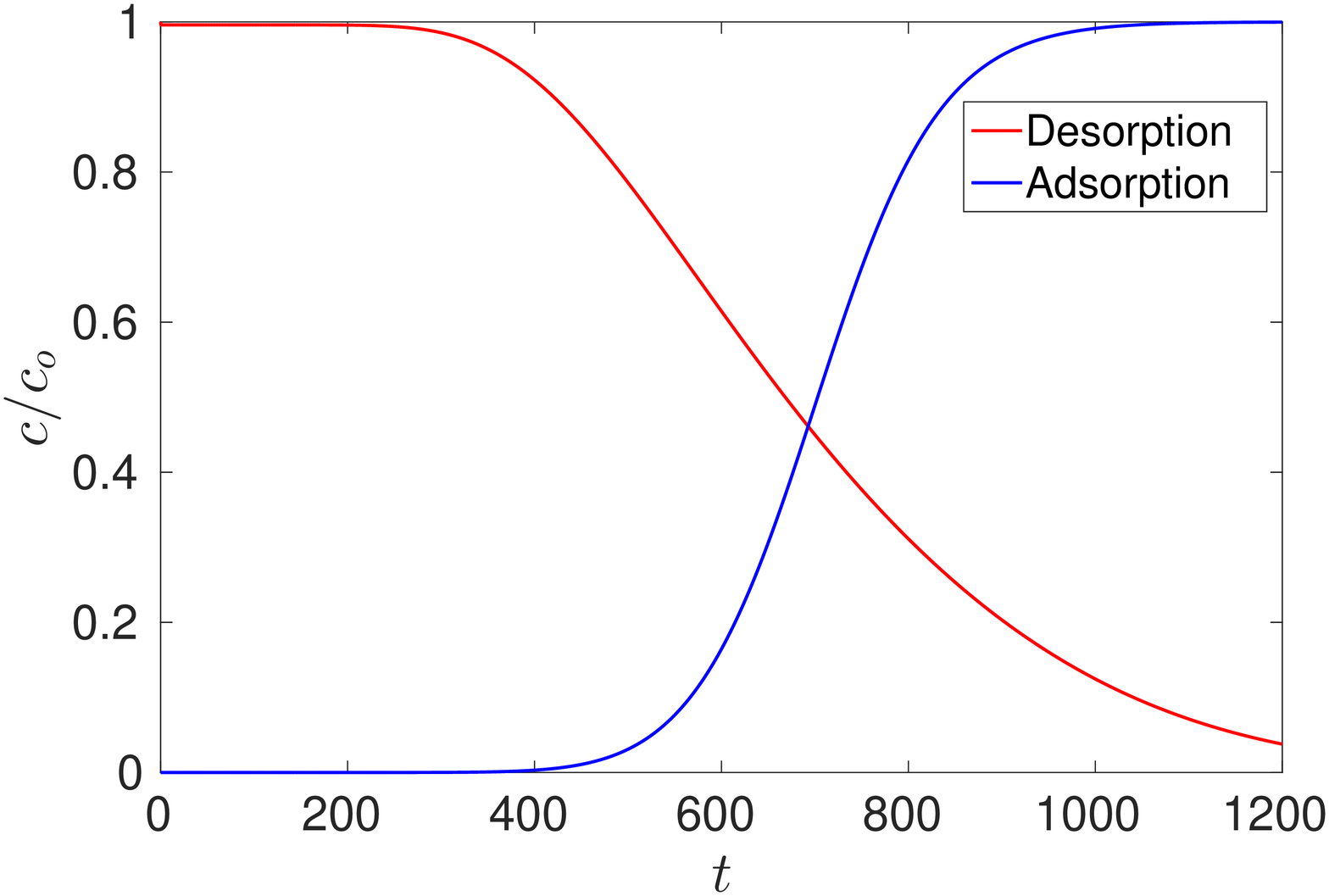}
\caption{Adsorption and Desorption breakthrough curve for single component}
\label{singlezadde}
\end{figure}

The breakthrough curve for desorption of single component by an inert fluid is shown in Fig~\ref{singlezdesorp}. Same values of the parameters given in Table1 are used for simulation whereas $q_o^*$ is given as,

\begin{equation}
q_o^{*}=\frac{q_mbc_o}{1+bc_o}=0.0092 \frac{kg~adsorbed}{kg~adsorbent}
\end{equation}
If we plot both adsorption and desorption breakthrough curve (see Fig~\ref{singlezadde}) on the same graph, then we can see that the curves are asymmetric for nonlinear isotherm. As the value of b increases,
asymmetry of the curves increases. But for linear isotherm curves are symmetric. For linear isotherm ordinate value of point of intersection between adsorption and desorption breakthrough curves is 0.5. But for nonlinear isotherm, it is always below 0.5. As the value of
b increases it dips further below 0.5.

\section{Multi-component adsorption breakthrough curve}

\paragraph{Mathematical modeling:}
Model principles and assumption will remain same for this case, but here governing equation
are solved for all the components. Binary component adsorption is considered
here. We consider a case where linear driving force of both the components
are equal with different isotherm properties. Governing equations are as follows:
\begin{eqnarray}
u \frac{\partial c_1}{\partial z}+\frac{\partial c_1}{\partial t}+\frac{1-\epsilon}{\epsilon}\rho_p\frac{\partial q_1}{\partial t}=0 \\
u \frac{\partial c_2}{\partial z}+\frac{\partial c_2}{\partial t}+\frac{1-\epsilon}{\epsilon}\rho_p\frac{\partial q_2}{\partial t}=0 \\
\frac{\partial q_1}{\partial t}=(LDF)_1\left(q_1^{*}-q_1\right) \\
\frac{\partial q_2}{\partial t}=(LDF)_2\left(q_2^{*}-q_2\right)
\end{eqnarray}

whereas adsorption isotherms are described by extended Langmuir isotherm,

\begin{eqnarray}
q_1^{*}=\frac{q_{m_1}b_1 c_1}{1+b_1 c_1+ b_2 c_2}\\
q_2^{*}=\frac{q_{m_2}b_2 c_2}{1+b_1 c_1+ b_2 c_2}
\label{multiso}
\end{eqnarray}

The following initial conditions are considered,
\begin{eqnarray}
\nonumber c_1=c_{10} &&  z=0, t=0 \\ \nonumber
c_2=c_{20} &&  z=0, t=0 \\ \nonumber
q_1=0 &&    0<z\leq L, t=0 \\ \nonumber
q_2=0 &&    0<z\leq L, t=0 \\ \nonumber
c_1=0 &&   0<z\leq L, t=0 \\ \nonumber
c_2=0 &&   0<z\leq L, t=0
\label{icsmult}
\end{eqnarray}
Boundary conditions for the problem is given as,
\begin{eqnarray}
\nonumber c_1=c_{10} &&  z=0,t>0 \\ \nonumber
c_2=c_{20} &&  z=0,t>0 
\label{bcsmult}
\end{eqnarray}

Governing equations are solved along with the adsorption isotherms and boundary conditions as well as
initial conditions to yield the breakthrough curve for two components competitive adsorption.

\paragraph{Results:}
\begin{table}
\caption*{Table 2}
\begin{center}
  \begin{tabular}{ |c|c|c|}
    \hline
    Parameter & Value \\ \hline
    $L$,Bed length,m & $0.3$ \\ \hline
    $\epsilon$ & $0.4$ \\ \hline
    $u$, Velocity ,$m/s$ & $0.01$ \\ \hline
    $(LDF)_1$ $(sec)^{-1}$ & 1.5 \\ \hline
    $(LDF)_2$ $(sec)^{-1}$ & 1.5 \\ \hline
    $b_1$ $m^3/kg$ & $0.4$ \\ \hline
    $b_2$ $m^3/kg$ & $0.3$ \\ \hline
    $q_{m_1}$ kg adsorbed/kg adsorbent & $0.04$ \\ \hline
    $q_{m_2}$ kg adsorbed/kg adsorbent & $0.03$ \\ \hline
    $c_{10}$, inlet conc. of the component $1$ in the feed , $kg/m^3$ & $0.75$ \\ \hline
    $c_{20}$, inlet conc. of the component $1$ in the feed , $kg/m^3$ & $0.5$ \\ \hline
    $\rho_p$ $kg/m^3$ & $800$\\ \hline
  
  \end{tabular}
  
  \label{parametermult}
\end{center}
\end{table}
\begin{figure}
\includegraphics[scale=.35]{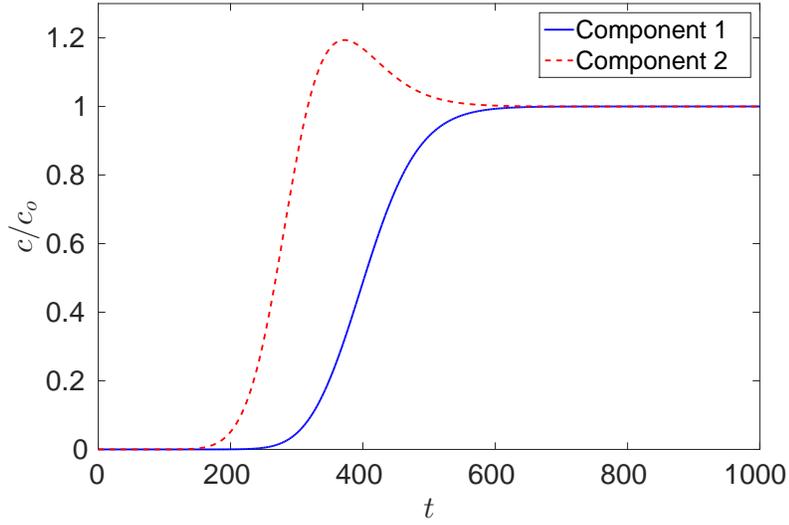}
\caption{Multicomponent adsorption breakthrough curve}
\label{multiadbreak}
\end{figure}

Value of the parameters for simulation is given in Table~2. And the breakthrough curves
obtained from simulation is shown in Fig~\ref{multiadbreak}. From the plot it is found that for one component
effluent concentration has exceeded feed concentration. This phenomenon is known as roll up
or roll over. For this case roll up is caused by the displacement of a weaker adsorbate by a
stronger one.

\section{Generic breakthrough curve}

In order to determine the breakthrough nature for adsorption process under different operating conditions we need to solve
the model equations for each case. But if we can capture what change in the breakthrough
curve is brought about by the change in the value of an operating variable, then it is possible
to say what would be the nature of a new breakthrough curve for the changed value of the
variable with reference to a known breakthrough curve without doing numerical simulation.
This can be done for different variables and then the results obtained for each case can be
clubbed together to accommodate the change in the breakthrough curve if all the variables
change simultaneously. So initially a parametric study is performed to find out the influence of different parameters on the breakthrough curve and the curves are properly rescaled to obtain a generic breakthrough curve. To accomplish this objective, the same
mathematical model as described earlier has been used to describe transient adsorption
process in a fixed bed adsorber for a single component adsorption onto a monolayer adsorbent. After nondimensionalization of model equations, it has been solved numerically
with proper boundary and initial conditions. Then a parametric study is done to generate the
the universal breakthrough curve.

\paragraph{Mathematical modeling}

Same model described earlier is used here. But here governing equations, boundary
conditions and initial conditions are nondimensionalized. Parameters used for
nondimensionalization and nondimensionalized equations, boundary conditions and initial
conditions are as follows,

Dimensionless parameters:

\begin{eqnarray}
\nonumber c^{\prime}=\frac{c}{c_0}; ~ t^{\prime}=\frac{15D_et}{R_p^2};~ z^{\prime}=\frac{15D_e}{R_p^2}\left(\frac{z}{u}\right)\\ \nonumber
\rho_{p}^{\prime}=\left(\frac{1-\epsilon}{\epsilon}\right)\frac{\rho_{p}}{c_0};~b^{\prime}=bc_0
\end{eqnarray}

This leads to the following nondimensionalized equations
\begin{eqnarray}
 \frac{\partial c^{\prime}}{\partial z^{\prime}}+\frac{\partial c^{\prime}}{\partial t^{\prime}}+\rho_{p}^{\prime}
\left(\frac{q_m b^{\prime} c^{\prime}}{1+b^{\prime} c^{\prime}}-q\right)=0 \\
\frac{\partial q}{\partial t^{\prime}}=  \left(\frac{q_m b^{\prime} c^{\prime}}{1+b^{\prime} c^{\prime}}-q\right)&&
\end{eqnarray}

Modified initial and boundary conditions are as follows ,

Initial conditions:

\begin{eqnarray}
\nonumber c^{\prime}=1 &&  z^{\prime}=0, t^{\prime}=0 \\ \nonumber
q=0 &&    0<z^{\prime}\leq \frac{15D_e}{R_p^2}\left(\frac{L}{u}\right), t^{\prime}=0 \\ 
c^{\prime}=0 &&    0<z^{\prime}\leq \frac{15D_e}{R_p^2}\left(\frac{L}{u}\right), t^{\prime}=0
\label{icsnonsingle}
\end{eqnarray}

Boundary conditions are given as;
\begin{eqnarray}
c^{\prime}=1 && z^{\prime}=0, t^{\prime}>0
\end{eqnarray}

\paragraph{Results and analysis}

\begin{figure}
\includegraphics[scale=.35]{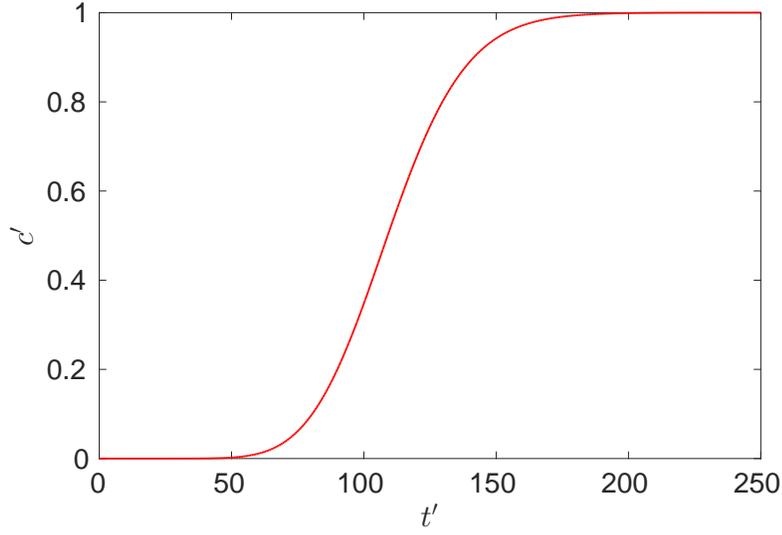}
\caption{Reference breakthrough curve}
\label{refbreak}
\end{figure}

\begin{table}\
\caption*{Table 3}
\begin{center}
  \begin{tabular}{ |c|c|c|}
    \hline
    Parameter & Value \\ \hline
    $L$,Bed length,m & $0.3$ \\ \hline
    $\epsilon$ & $0.4$ \\ \hline
    $u$, Velocity ,$m/s$ & $0.005$ \\ \hline
    $(LDF)$ $(sec)^{-1}$ & 0.5 \\ \hline
    $b$ $m^3/kg$ & $0.1$ \\ \hline
    $q_{m}$ kg adsorbed/kg adsorbent & $0.02$ \\ \hline
    $c_0$, inlet conc. of the adsorbate in the feed , $kg/m^3$ & $1.5$ \\ \hline
    $\rho_p$ $kg/m^3$ & $1100$\\ \hline
  
  \end{tabular}
  \label{parametermbreak}
\end{center}
\end{table}
At the outset the model equations are solved for some chosen values of different parameters.
Now as breakthrough curve is highly sensitive to
different parameters like bed length, superficial velocity, diffusivity, adsorbent particle
radius, isotherm properties of adsorbent etc., we will initially see how breakthrough pattern
changes with change in the value of these parameters separately and a function is developed
in each case to predict breakthrough pattern with reasonable accuracy for different scenarios.
Then all the parameters are varied simultaneously and accordingly rules are set up to predict
the nature of the breakthrough for different cases. First a reference breakthrough curve is produced and values of the
parameters for the reference breakthrough curve are given in Table~3. The reference breakthroiugh curve is shown in Fig~\ref{refbreak}.

\begin{figure}
\includegraphics[scale=.35]{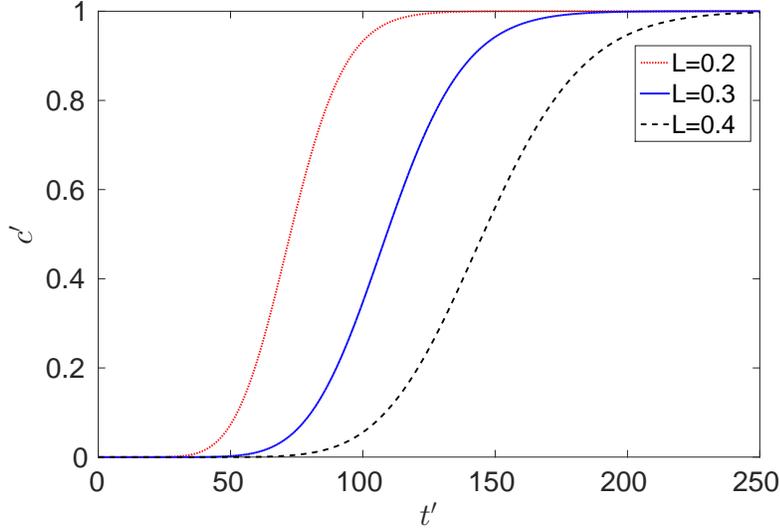}
\caption{Effect of bed length on the breakthrough curve}
\label{length}
\end{figure}

\paragraph{Effect of bed length, L} 

Fig~\ref{length} shows how the bed length affects the breakthrough pattern. It is seen from the plot
that as the bed length increases breakthrough point also shifts towards right along the time
scale. Smaller bed length corresponds to lesser amount of adsorbent. Consequently a smaller
capacity for the bed to adsorb and the bed gets saturated in less time.

\begin{figure}
\includegraphics[scale=.3]{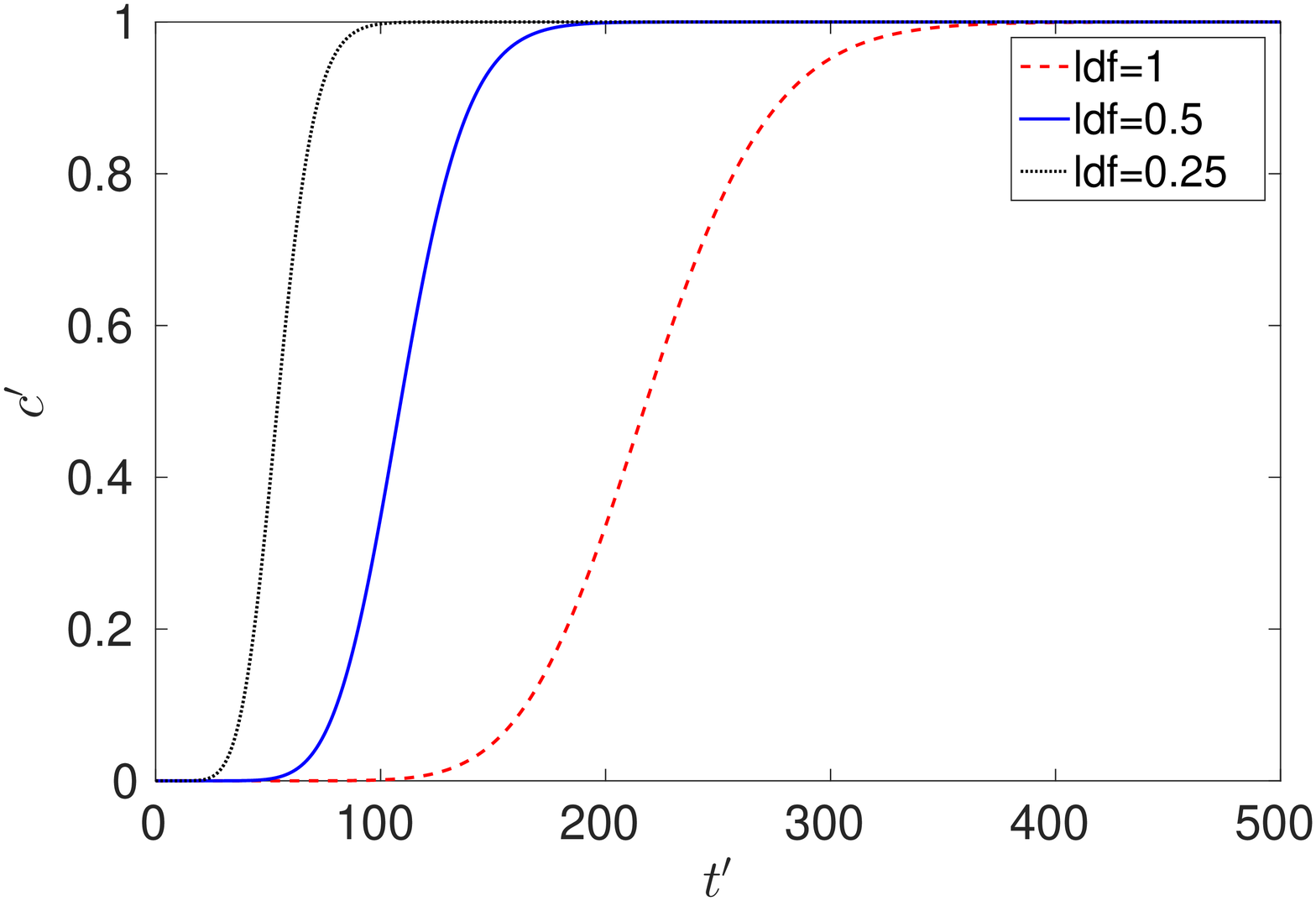}
\includegraphics[scale=.3]{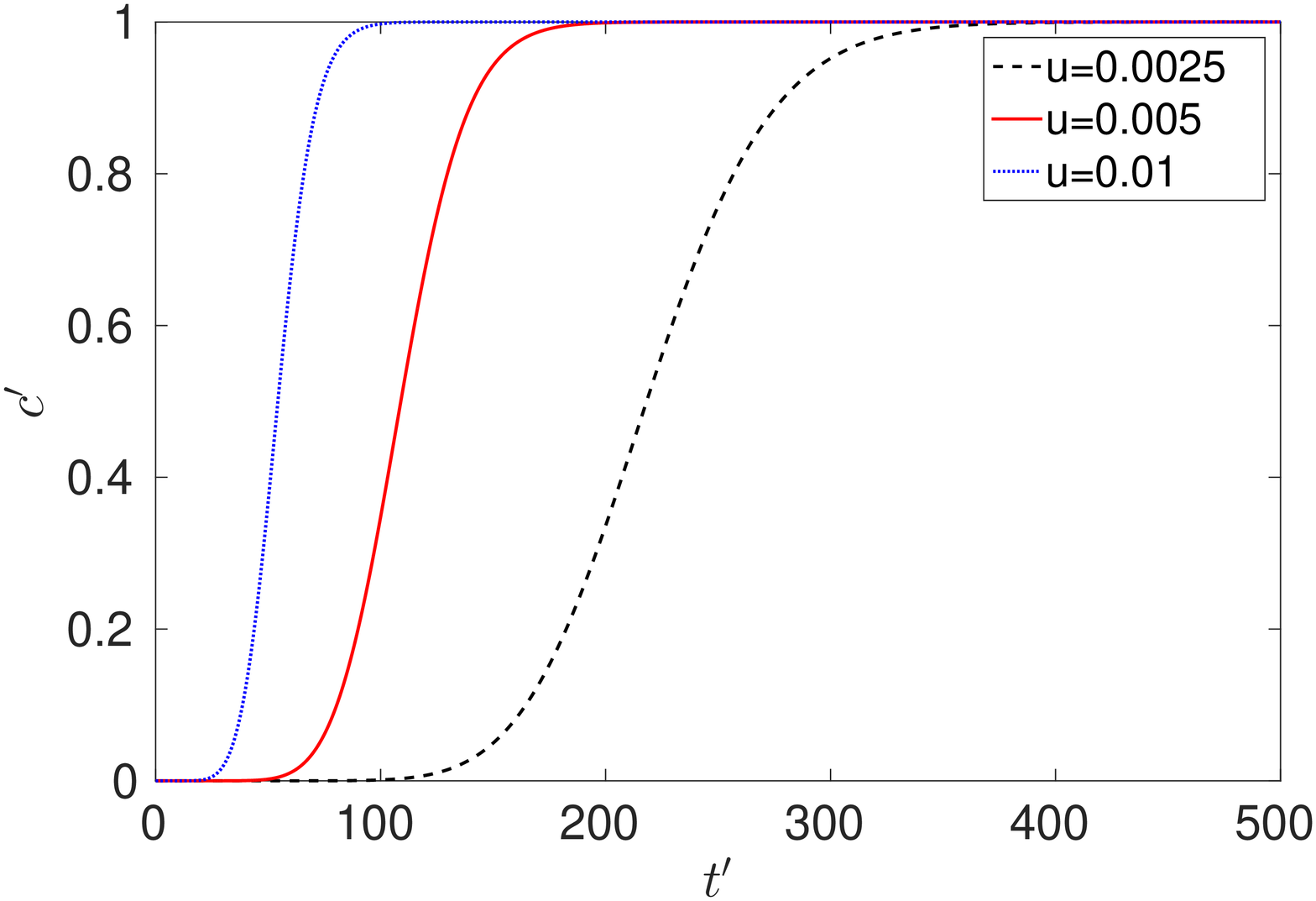}
\caption{Upper panel: Effect of LDF $\left(15D_e/Rp^2\right)$, Lower panel: Effect of velocity on the breakthrough curve}
\label{ldf}
\end{figure}

\begin{figure}
\includegraphics[scale=.3]{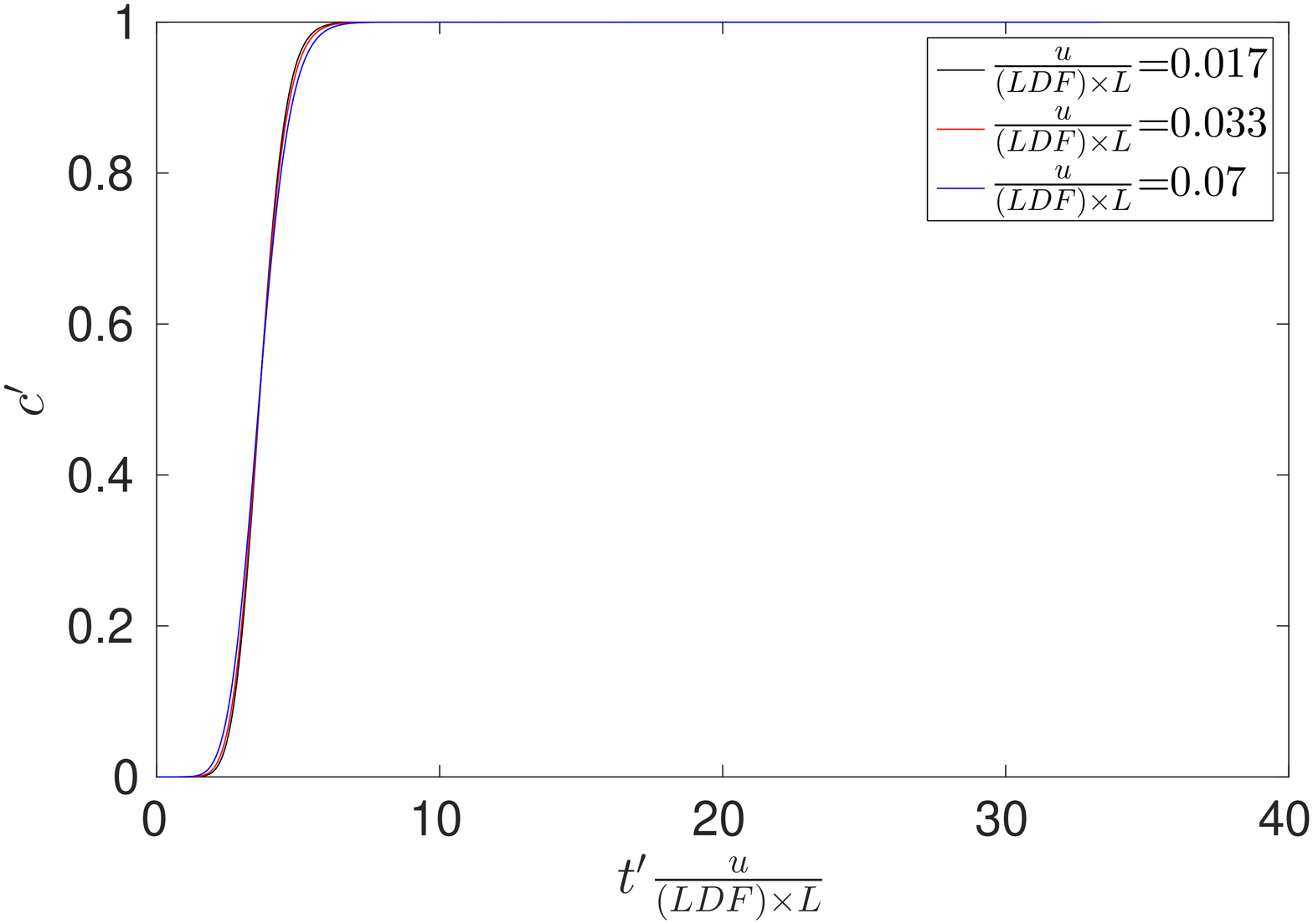}
\caption{Collapse of breakthrough curve for different value of $\frac{u}{LDF \times L}$  }
\label{clubbed}
\end{figure}
\paragraph{Effect of LDF $\left(15D_e/Rp^2\right)$}
Fig~\ref{ldf} upper panel shows how LDF affects breakthrough nature. It is seen from the plot that as LDF
increases breakthrough point shifts towards right along the time scale. As LDF is the measure
of the uptake rate of the adsorbent particle, when its value is lower the bed gets saturated
faster than the higher value of LDF.

\begin{figure}
\includegraphics[scale=.3]{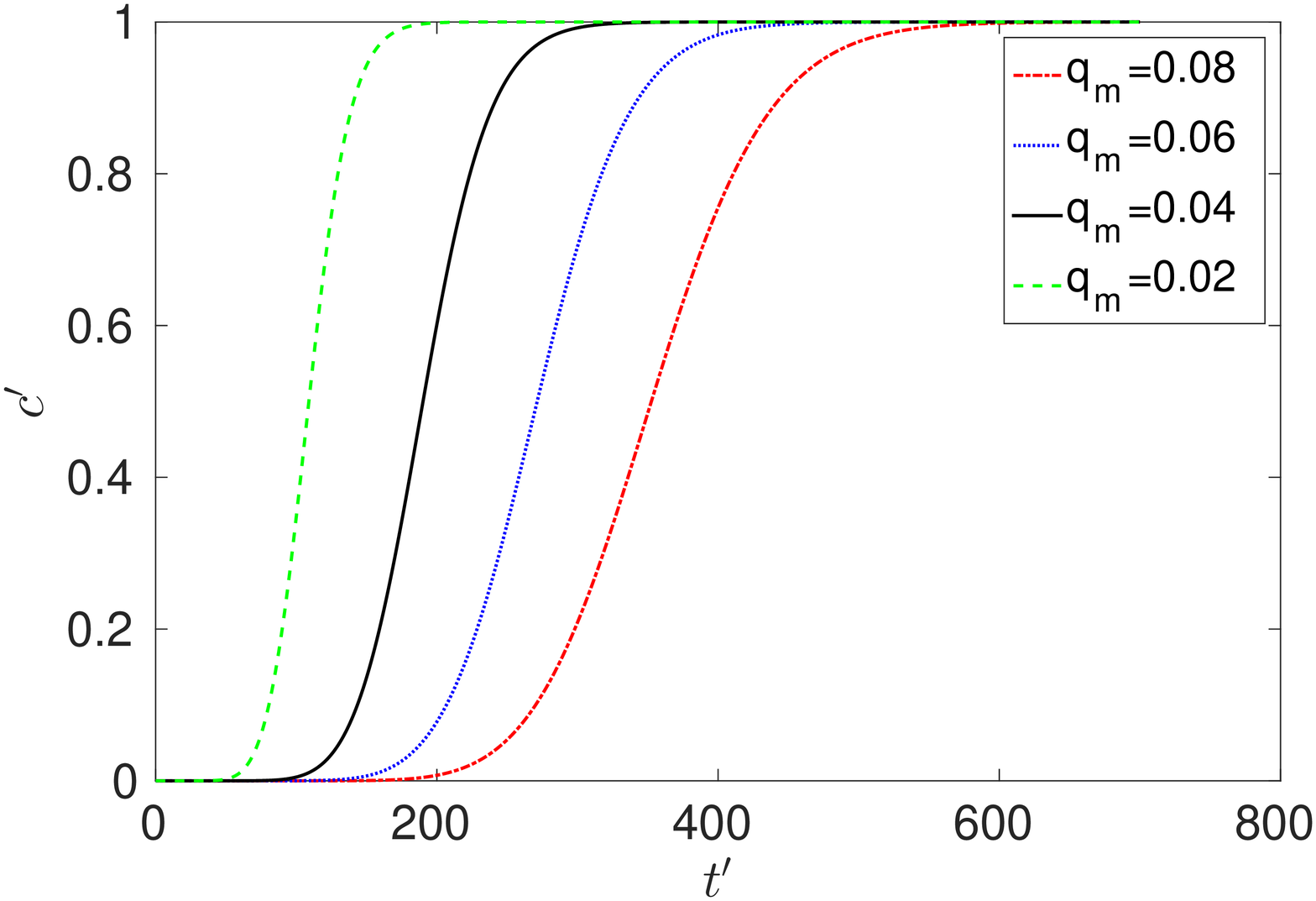}
\caption{Effect of maximum adsorption capacity, $q_m$ on the breakthrough curve}
\label{qm}
\end{figure}
\begin{figure}
\includegraphics[scale=.3]{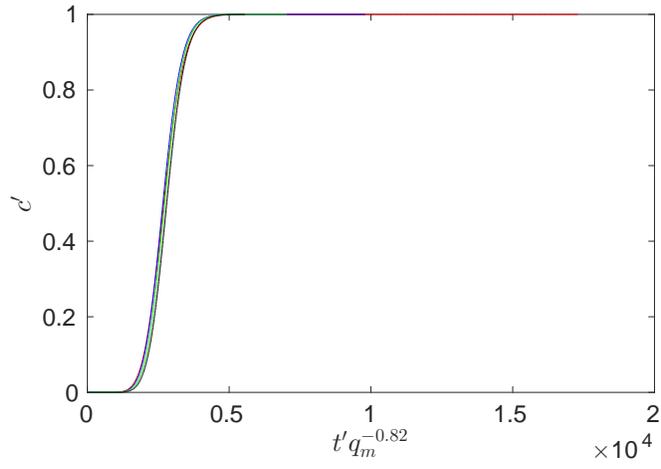}
\caption{Data collapse of breakthrough curves for different values of maximum adsorption capacity, $q_m$}
\label{qmscale}
\end{figure}

\begin{figure}
\includegraphics[scale=.3]{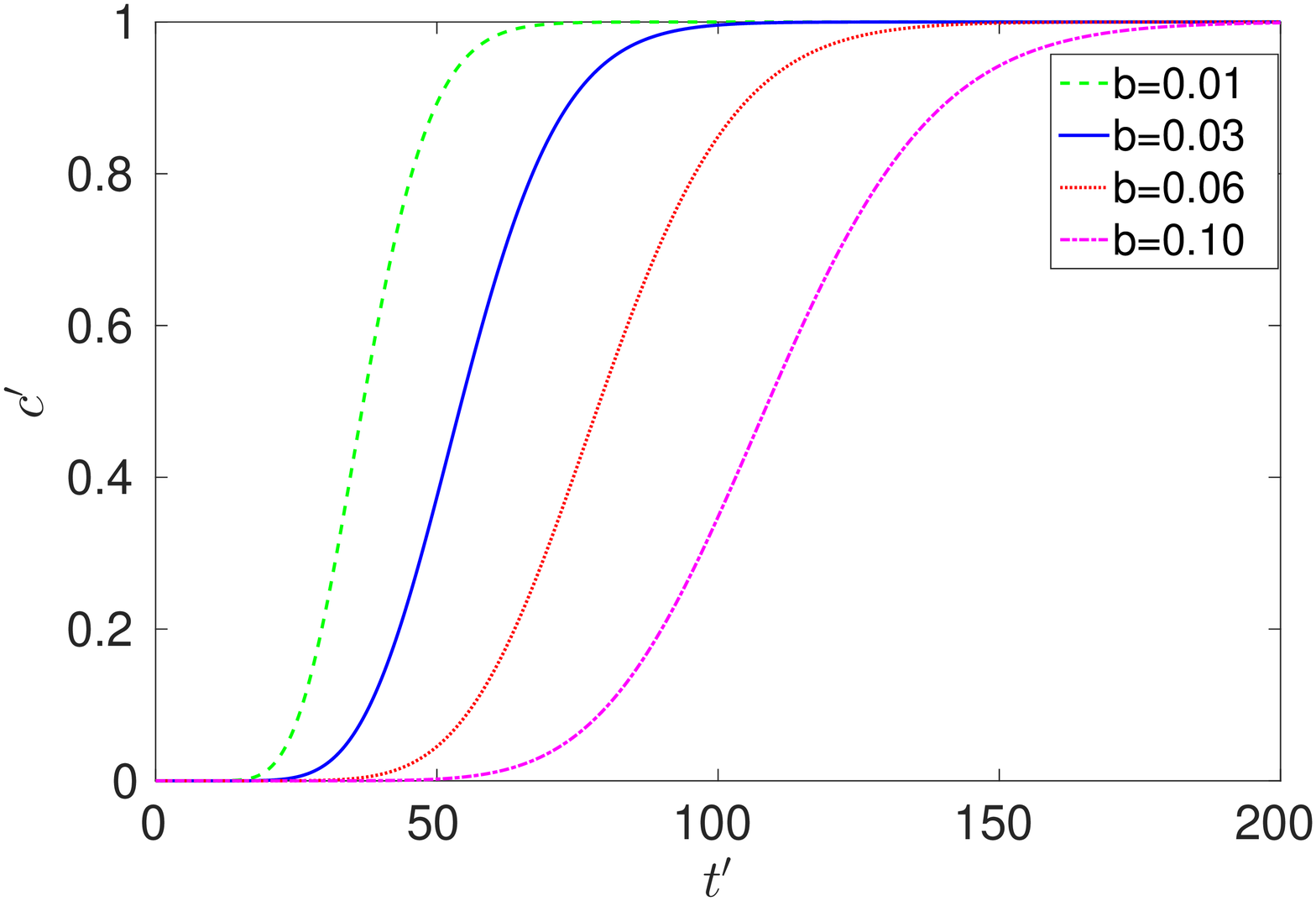}
\caption{Effect of Langmuir isotherm constant, $b$ on the breakthrough curve}
\label{bd}
\end{figure}

\begin{figure}
\includegraphics[scale=.3]{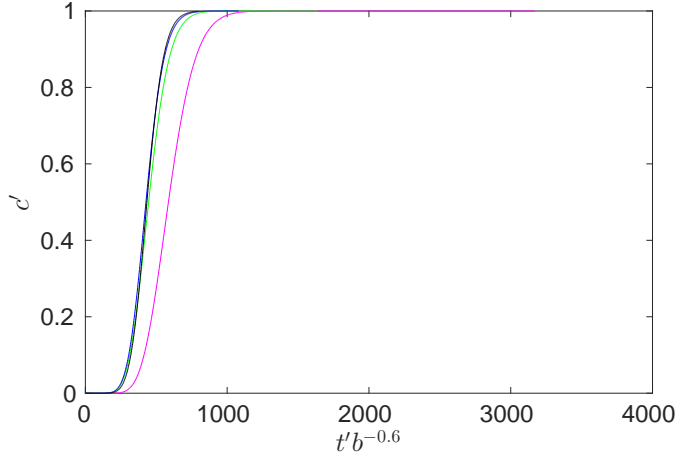}
\caption{Data collapse of breakthrough curves for different values of Langmuir isotherm constant, $b$}
\label{bdscale}
\end{figure}

\paragraph{Effect of velocity,$u$}
Fig~\ref{ldf} lower panel shows that as the velocity increases breakthrough point shifts towards the left and also curve becomes much steeper. This is because of the residence time of the solute in the column, which is not long enough for adsorption equilibrium to be reached at high velocity. So at high velocity the adsorbate solution leaves the column before equilibrium occurs. Furthermore, a fixed saturation capacity of bed based on the same driving force gives rise to a shorter time for saturation at higher velocity.

The effect of velocity, linear driving force constant and the bed length can be clubbed into a non-dimensional number $\frac{u}{LDF \times L}$ and when this number is multiplied with the dimensionless time scale, then the breakthrough curve collapses onto a single curve for different values of this number i.e different values of $L,u$ and $LDF$. The same is shown in Fig.~\ref{clubbed}

\paragraph{Effect of maximum adsorption capacity ,$q_m$}
Fig~\ref{qm} shows that as $q_m$ increases breakthrough point shifts towards the right of the
dimensionless time scale. As $q_m$ increases, equilibrium adsorption capacity of the adsorption
particle increases. So it takes more time for the bed to get saturated for higher value of $q_m$. The breakthrough curves for different $q_m$ can be collapsed onto a single curve for the values considered if the dimensionless time is multiplied with $q_m^{-0.82}$. Such scaling is shown in Fig~\ref{qmscale}.

\paragraph{Effect of Langmuir isotherm constant,$b$}
Fig~\ref{bd} shows that as value of $b$ increases, breakthrough point shifts towards the right of the
dimensionless time scale. As the value of $b$ increases, equilibrium adsorption
capacity of the adsorbent particle increases. So it takes longer time for the bed to get saturated
at higher value of $b$. The breakthrough curves for different $b$ can be collapsed onto a master curve for the values considered if the dimensionless time is multiplied with $b^{-0.6}$ and re-plotted. The scaling is shown Fig~\ref{bdscale}. The scaling seems to deviate as the value of $b$ increases.

\section{Conclusions}
The present work is a building block to the understanding of more complicated pressure
swing adsorption process. Basic principles of adsorption and adsorption process are
understood through a simple mathematical model. Solution of the model yielded breakthrough curves for single
component and multi-component adsorption. Also fluid phase concentration profile of
adsorbate is obtained from simulation. Breakthrough curve for desorption of a saturated bed
by an inert fluid is also obtained after solving model equations with different initial and
boundary condition compared to adsorption. A detailed parametric study is performed to get an insight on the effects of different influencing and crucial parameters like bed length, velocity,
diffusivity, particle radius and isotherm properties on the nature of the breakthrough curve for adsorption process. These
results can be very useful in designing of adsorption columns. Analysis of these results led to
the development of generic breakthrough curve that will enable us to tell the nature of
breakthrough curve for different process parameters without recourse to the numerical
simulation for a single component monolayer adsorption. Thus we can save on computational/experimental
time. Future work will involve more realistic mathematical model and determining generic breakthrough curve for the same.

\section{References}

\bibliography{mybibfile}

\end{document}